\documentclass[a4paper]{article}
\usepackage{cite}
\usepackage{graphicx}
\usepackage{tabularx}
\usepackage{amsmath}
\usepackage{array}
\usepackage[nameinlink, capitalise, noabbrev]{cleveref}

\usepackage{INTERSPEECH2020}

\title{Feature Fusion Strategies for End-to-End Evaluation\\ of Cognitive Behavior Therapy Sessions}
\name{Zhuohao Chen$^1$, Nikolaos Flemotomos$^1$, Victor Ardulov$^1$, Torrey A. Creed$^2$,\\ Zac E. Imel$^3$, David C. Atkins$^4$, Shrikanth Narayanan$^1$}
\address{
  $^1$Signal Analysis and Interpretation Lab, University of Southern California, Los Angeles, CA, USA\\
  $^2$Department of Psychiatry, University of Pennsylvania, Philadelphia, PA, USA\\
  $^3$Department of Educational Psychology, University of Utah, Salt Lake City, UT, USA\\
  $^4$Department of Psychiatry and Behavioral Sciences, University of Washington, Seattle, WA, USA}
\email{$^1$sail.usc.edu,$^2$tcreed@pennmedicine.upenn.edu,$^3$zac.imel@utah.edu,$^4$datkins@u.washington.edu}

\begin{document}

\maketitle
\begin{abstract}
  Cognitive Behavioral Therapy (CBT) is a goal-oriented psychotherapy for mental health concerns implemented  
  in a conversational setting with broad empirical support for its effectiveness across a range of presenting problems and client populations. The quality of a CBT session is typically assessed by trained human raters who manually assign pre-defined session-level behavioral codes. In this paper, we develop an end-to-end 
  pipeline that 
  converts 
  speech audio to diarized and transcribed text and extracts linguistic features to code the CBT sessions automatically. We investigate both word-level 
  and utterance-level features and propose feature fusion strategies 
  to combine them. The utterance level features include dialog act tags as well as behavioral codes drawn from another well-known talk psychotherapy called Motivational Interviewing (MI). We propose a novel method to  
 augment the word-based features with the utterance level tags for subsequent CBT code estimation.
  Experiments show that our new fusion strategy outperforms all the studied features, both when used individually and when fused by direct concatenation. We also find that incorporating a sentence segmentation module can further improve the overall system given the preponderance of multi-utterance conversational turns in CBT sessions.
  
\end{abstract}
\noindent\textbf{Index Terms}: cognitive behavioral therapy, end-to-end evaluation, feature fusion strategies

\section{Introduction}

In psychotherapy assessment, the quality of a session is generally evaluated through the process of behavioral coding in which experts manually identify and annotate behaviors of the participants \cite{bakeman2000behavioral}. However, this procedure is time-consuming, which makes it resource-heavy in terms of human capital and therefore often unfeasible in most treatment contexts. In recent years, researchers have developed automated behavioral coding algorithms using speech and language features for several clinical domains such as addiction counseling \cite{xiao2016behavioral}, post-traumatic stress disorder (PTSD) care \cite{shiner2012automated} and autism diagnosis \cite{bone2016use}. The work in \cite{hirsch2018s} even presents an automated evaluation Motivational Interviewing (MI) \cite{miller2012motivational} system which avoids both manual annotation and transcription. Some studies extend to multimodal approaches which also take non-lexical characteristics into consideration \cite{ardulov2018multimodal, singla2018using, chen2019improving}. 

Cognitive Behavioral Therapy (CBT) is evidence-based psychotherapy predicated on the cognitive model which involves shifts in the patient’s thinking and behavioral patterns~\cite{beck2011cognitive}. In a CBT session, the therapist guides individuals to identify goals and develop new skills to overcome obstacles that interfere with goal attainment. As a common type of talk therapy, CBT has been developed for many decades and has become an effective treatment for a wide range of mental health conditions \cite{hofmann2012efficacy}. Extending upon this strong evidence base, recent research has explored whether combining CBT with other evidence-based psychotherapies might potentiate treatment outcome. For example, studies indicate that adding MI as an adjunct to CBT may benefit patients by increasing motivation for and commitment to the intervention \cite{westra2006preparing, randall2017motivational}.


One of the early computational behavioral coding efforts for CBT is
found in \cite{flemotomos2018language} which employed an end-to-end evaluation pipeline that overcomes the need of manual transcription and coding. This work formulated
the CBT session quality evaluation as a classification task and compared the performance of various lexical features.

In this paper, we develop a new automated approach to assess CBT session quality. 
Specifically, we utilize MI data to extract utterance-level features due to the similarities between MI and CBT and propose a novel fusion strategy. 
We experiment on both manual transcripts and automatically derived ones to show the superiority of the new fusion approach the and robustness of our automated evaluation system.

\section{Datasets}

The CBT data, with accompanying audio-recorded sessions, used in this work come from the Beck Community Initiative~
\cite{creed2016implementation}, a large-scale public-academic partnership to implement CBT in community mental health settings.
The CBT quality is evaluated by the session-level behavioral codes based on Cognitive Therapy Rating Scale (CTRS) \cite{young1980cognitive}. Each session receives 11 codes scored on a 7 point Likert scale ranging from 0 (poor) to 6 (excellent) for each evaluated dimension.
We also compute the total CTRS by summing up the scores as an overall measurement of the quality of a session. Raters were doctoral-level experts who were required to demonstrate calibration prior to coding process to prevent rater drift, which resulted in high inter-rater reliability for the CTRS total score (ICC = 0.84).

In this paper, we use 225 coded CBT sessions for experiments which include the 92 sessions used in \cite{flemotomos2018language} with the highest and lowest total CTRS and 133 additional sessions to balance the distribution. Each session considered is a dyadic conversation between a therapist and a patient. We manually transcribed the sessions including information about talk turns, speaker roles, and punctuation. The sessions were recorded at a 16kHz sampling rate
and their lengths range from 10 to 90 minutes. We binarized the CTRS codes by assigning codes greater or equal to 4 as ``high" and less than 4 as ``low" since 4 is the primary anchor 
indicating the skill is fully present, but still with room for improvement \cite{young1980cognitive}. The threshold indicative of CBT competence on the total CTRS is 40 \cite{vallis1986cognitive}. The descriptions and the label distributions of the codes are shown in Table \ref{tab:labels}.

\begin{table}[htb]
  \caption{CBT behavior codes defined by the CTRS manual}
  \label{tab:labels}
  \centering
  \resizebox{0.37\textwidth}{!}{\begin{tabular}{lll}
    \toprule
    Abbr.      & CTRS Code     & low/high           \\
    \midrule
    ag  & agenda                                      &  131/94                                 \\
    at & \begin{tabular}[l]{@{}l@{}}application of cognitive-\\ behavioral techniques\end{tabular} & 150/75                                \\
    co  & collaboration                               & 111/114                                  \\
    fb  & feedback                                    &  150/75                                 \\
    gd  & guided discovery                            & 146/79                                  \\
    hw  &  homework                                   & 165/60                                 \\
    ip  & interpersonal effectiveness                 & 47/178                                  \\
    cb  & \begin{tabular}[l]{@{}l@{}}focusing on key cognitions\\ and behaviors\end{tabular}    & 122/103                                 \\
    pt  & pacing and efficient use of time            & 135/90                                  \\
    sc  & strategy for change                         & 126/99                                  \\
    un  & understanding                               &  123/102                                \\
    \midrule
    total                 & total score              & 134/91                                   \\
    \bottomrule
  \end{tabular}}
\end{table}

\section{Approach}

Our end-to-end evaluation approach includes two stages. In the first stage, we took the session recordings as inputs and used a speech processing pipeline to substitute manual transcription. In the second stage, we 
extracted the linguistic features from the therapist's transcripts to predict the binarized label of each code. The classification tasks are performed by a linear Support Vector Machine (SVM) with sample weights inversely proportional to their class frequencies.

\subsection{Speech Processing Pipeline}\label{sec:3-1}

\begin{figure}[htb]
  \centering
  \includegraphics[width=7.0cm]{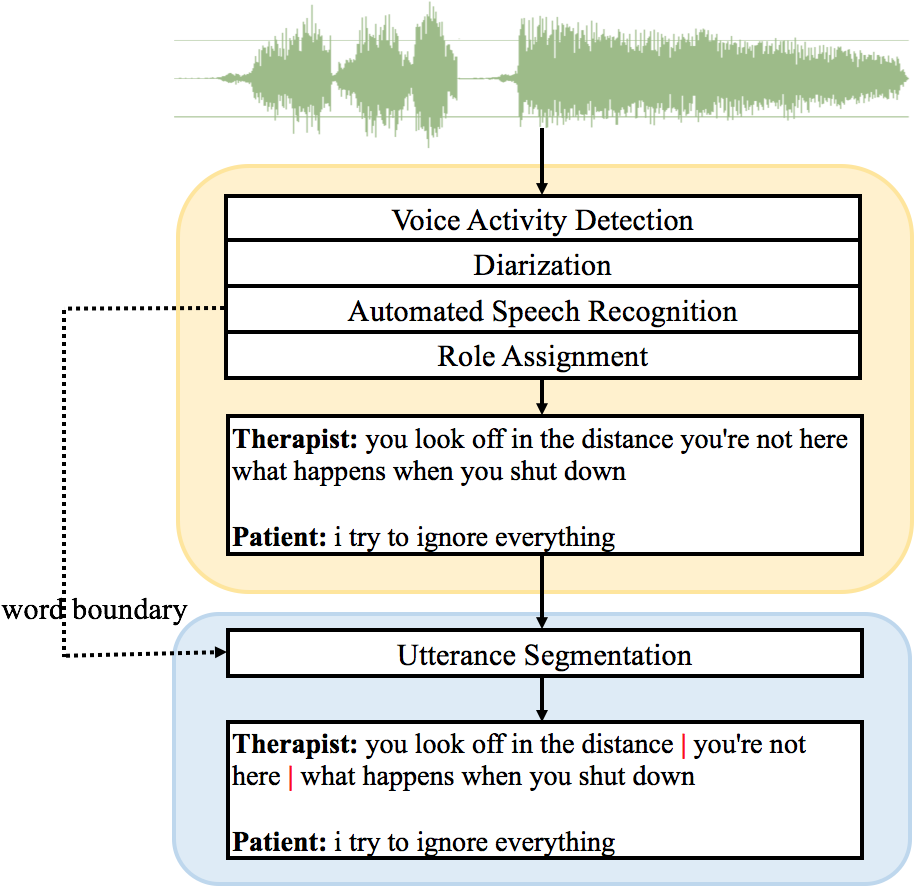}
  \caption{Session Decoding Pipeline}
  \label{fig:pipe}
\end{figure}

To automatically transcribe the recorded sessions we adopted the speech pipeline described in \cite{martinez2019identifying} consisting of 
Voice Activity Detection (VAD), diarization, Automatic Speech Recognition (ASR) and role assignment presented by the yellow box in Fig.~\ref{fig:pipe}. The diarization error rate (including VAD errors) and ASR word error rate for the transcribed sessions are 21.47\% and 44.01\%, respectively. 
The role assignment module in \cite{martinez2019identifying} is trained to distinguish the therapist and patient in a counseling session and the annotation accuracy for the transcribed CBT sessions is 100\% (225/225).


As shown in Fig.~\ref{fig:pipe}, the output of the speech pipeline in the yellow box is at the turn level without any punctuation. There might be multiple utterances within a turn, something which potentially 
affects the quality of utterance-level lexical features, and the subsequent behavioral coding. Thus, we implemented an utterance segmentation module at the end of the pipeline. We made use of the word boundaries to split the text 
whenever the pause between consecutive words is more than 2 seconds, and then segmented the transcripts into utterances. The package we applied for utterance segmentation is an open source tool called ``DeepSegment" \cite{deepsegment}. It employs a bi-LSTM layer followed by a Conditional Random Field (CRF) \cite{lafferty2001conditional}. The model is trained using the Tatoeba corpus \cite{ho2016tatoeba} and it achieves an F1 score of 0.7364 on the transcribed sessions.

\subsection{Baseline Mid-level Features}\label{sec:3-2}

We extract a number of different mid-level features from the transcribed text. The first set includes the term frequency - inverse document frequency (tf-idf) \cite{dillon1983introduction} transform of n-grams, while the second focuses on estimated Dialog Acts (DA) \cite{stolcke2000dialogue}. The tf-idfs and DAs were reported to achieve the best overall performance among the interpretable features in \cite{flemotomos2018language}. While these aforementioned features are based on general spoken dialog characteristics, the third feature set considered here is inspired by utterance level codes drawn from MI. Under the hypothesis that there are shared characteristics between MI and CBT (based on prior work that has reported using MI techniques as a way to facilitate CBT treatments \cite{westra2006preparing, randall2017motivational}), we experimentally investigate the usefulness of MI based ``features" in contributing to the quality assessment of CBT. 
 
 We extract all these features for the therapist side of the conversation only, because, as reported in \cite{flemotomos2018language}, they perform robustly 
 for the task of behavioral coding, and further fusing features of the two roles (i.e., therapist and patient) does not lead to substantial improvements.
 We compute the tf-idfs over unigrams. 
We additionally tag each utterance in a CBT session by one DA from the 7-class scheme described in Table~\ref{tab:DA_MC}. We used a linear chain CRF model trained on the Switchboard-DAMSL dataset \cite{can2015dialog} which achieves 84.78\% accuracy of the in-domain test set.
For the DA-based feature representation 
we 1) count the utterances coded with each DA and normalize the counts with respect to the total number of utterances in each session; 2) count the words in the utterances tagged by each DA and normalize the count with respect to the total number of words in each session. 
Concatenating the two sets of normalized features, we get a DA feature set of 7 $\times$ 2 = 14 dimensions that outperforms the individual use of either of those sets.

To capture MI-like approaches used within a CBT session, we use specific utterance-level representations that describe MI relevant behaviors in the conversation. In particular, we employ the set of Motivational Interviewing Skills \cite{miller2003manual} codes
described in \cite{xiao2016behavioral} and summarized in Table~\ref{tab:DA_MC}. We cluster `RES' and `REC' into one class `RE' since they are domain-specific in distribution \cite{chen2020label} and easily confused with each other. 
We extract the MI relevant behavioral codes (MC, henceforth) the same way as in \cite{chen2019improving} which uses a neural architecture stacking an embedding layer, a bi-LSTM with attention layer and a dense layer. We train the model on the MI corpus from \cite{atkins2014scaling, baer2009agency} with train/validation/test split equal to 3/1/1 and the classification accuracy on the MI test set is 81.1\%. The final MC-based feature representation is the same as the DA-based described previously. As observed in Table~\ref{tab:DA_MC}, the main difference between the DAs and the MCs is that the former focus on the function of the dialog structure, while the latter emphasize on the critical and causal elements deemed useful in the psychotherapy.


The tf-idfs are computed with regards to the occurrence of words in the sessions while the DAs and MCs are both annotations extracted at the utterance level. 
On this basis, we group the basic features into word-level features (tf-idfs) and utterance-level features (DAs, MCs).

\begin{table}[htb]
\centering
\caption{Details of DA and MC}
  \label{tab:DA_MC}
\resizebox{0.35\textwidth}{!}{\begin{tabular}{|c|c|}
\hline
\begin{tabular}[c]{@{}c@{}}Coding\\ Schemes\end{tabular} & Codes                                                                                                                                                                                        \\ \hline
DA                                                       & \begin{tabular}[c]{@{}c@{}}Question, Statement, Agreement, Other\\ Appreciation, Incomplete, Backchannel\end{tabular}                                                                    \\ \hline
MC                                                       & \begin{tabular}[c]{@{}c@{}}Facility (FA), Giving Information (GI),\\ Reflection (RE), Closed Question (QUC),\\ Open Question (QUO), MI Adherent (MIA), \\ MI Non-Adherent (MIN)\end{tabular} \\ \hline
\end{tabular}}
\end{table}

\section{Feature Fusion Strategies}

In this section, we discuss two feature fusion methods for combining the word-level and the utterance-level features.

\subsection{Fusion by Concatenation}

The first fusion approach is straightforward, namely concatenation of the different feature sets. The hypothesis here is that the fused feature sets are complementary to each other so that they jointly carry richer information. 
Herein we combine the word-level feature tf-idfs with each of the utterance-level features (DAs, MCs) and denote the fused feature sets as tf-idfs + DAs and tf-idfs + MCs, respectively.

\subsection{Augmenting Words with Utterance Tags}

When we compute word-level features like tf-idfs and bag of words, contextual information is ignored. 
For example, the word ``homework” (an important element within CBT) in a question may denote that the therapist is checking if the patient has completed the given assignment, while in a reflection it might imply that the therapist is describing/confirming the assignment to/with the patient. The distribution of (just the) word ``homework" helps us evaluate how well a therapist incorporates the use of homework relevant to CBT. However, to incorporate the context in which they are used, we propose a fusion strategy of augmenting words with utterance level information.

\begin{figure}[htb]
  \centering
  \includegraphics[width=7.1cm]{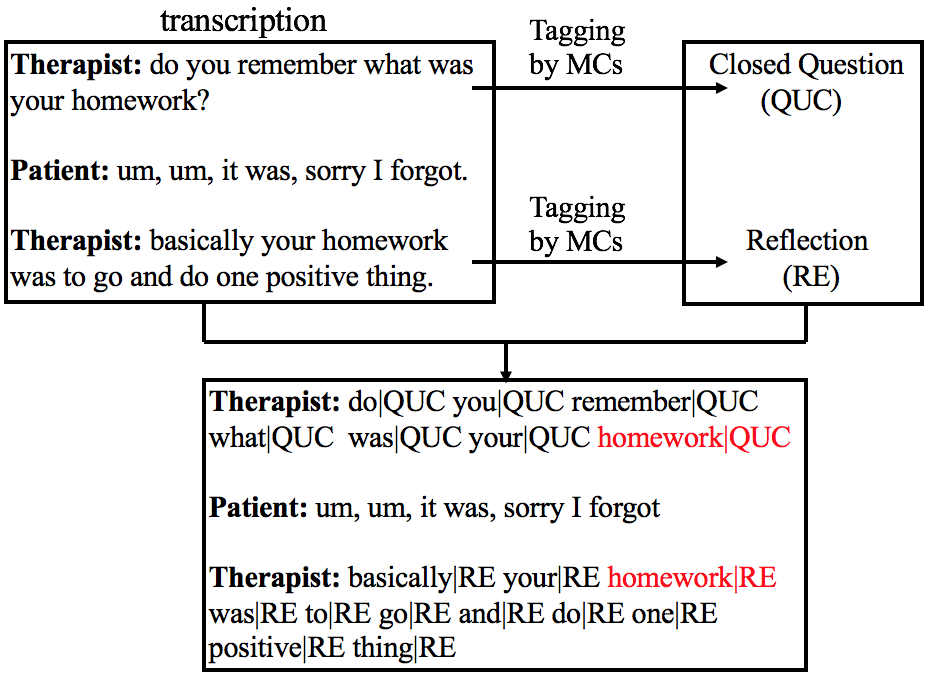}
  \caption{Word Augmentation}
  \label{fig:segment}
\end{figure}

We show an example of the word augmentation we propose using MCs in Fig.~\ref{fig:segment}. We first tag the therapist's utterances  by the model trained in \Cref{sec:3-2} and then pad the words with the label of the utterance they belong to. In Fig.~\ref{fig:segment} the augmented tokens ``homework$|$QUC" and ``homework$|$RE" are 
viewed as different words for further analysis. Finally, we extract the tf-idfs based on the augmented words of the therapist to obtain the fused features.

Similar to the previously-mentioned feature concatenation method, we fuse the augmented tf-idfs with each of the DAs and MCs  
and 
denote the fused feature sets as DA-tf-idfs and MC-tf-idfs, respectively.

\section{Experiments and Results}

In this section, we describe experiments on both the manual and the automatically derived transcripts of the CBT sessions. 

We compute the tf-idfs, DA-tf-idfs and MC-tf-idfs using the TfidfVectorizer from the scikit-learn Python module \cite{pedregosa2011scikit}. We set the parameters max\_df=0.95 and min\_df=0.05 to ignore terms that appear in more than 95\% or less than 5\% of the documents 
and select the K best features based on cross-validation on the total CTRS using a univariate F-test. All the feature sets are z-normalized before being fed into the linear SVM classifier. A 5-fold cross-validation is conducted to report the F1 score of each CTRS code and the total CTRS. The F1 scores are computed according to the total number of 
true and false positives over the folds \cite{forman2010apples}. 

\subsection{Results on Manual Transcripts}

The results 
of the classification task on the manual transcripts are presented in Table \ref{tab:manual}. 
From the reported results we find that the code `ip' (interpersonal effectiveness) -- which has the most imbalanced label distribution -- always has the lowest F1 score. Among the basic feature sets, the tf-idfs achieve a 
substantially better performance compared to either DAs or MCs, which indicates that these utterance-level features cannot fully capture CBT-relevant information contained in the word-level features.

Next we look at the direct fusion results of word level and utterance level features. By comparing the results of the tf-idfs with tf-idfs + DAs and tf-idfs + MCs we conclude that directly concatenating the tf-idfs with utterance-level features does not lead to substantial improvements,
confirming a similar conclusion drawn in \cite{flemotomos2018language}. Finally, we consider the proposed alternative fusion strategy. 
The performance of the DA-tf-idfs and MC-tf-idfs demonstrates that applying the new proposed fusion strategy to augment the words with the utterance tags, by either DAs or MCs, results in a better CBT relevant code prediction performance.
Especially the MC-tf-idfs -- which yield the best results among all the features sets -- significantly improve the F1 score of the total CTRS and averaged F1 score over tf-idfs (with $p$-value $<$ 0.05 based on the combined 5$\times$2cv F test \cite{alpaydm1999combined}).

It is interesting to point out that the MCs always lead to better performance compared to DAs, no matter whether we try to predict the CTRS codes by the basic feature set or after fusing with the tf-idfs.
This indicates that the behavioral codes defined in MI might exploit more useful therapy-relevant information, by encoding not only structural characteristics of the conversation, but also more psychotherapy-based cues. This also underscores the potential for transfer learning between MI and CBT (and perhaps other domains); some initial insights along these lines are provided in \cite{Gibson2019Multi-labelMulti-taskDeepLearning}.

\begin{table}[htb]
\caption{\label{tab:manual} F1 scores of the tasks on the manual transcripts.}
\centering
\resizebox{0.44\textwidth}{!}{\begin{tabular}{c|ccccccc}
\hline
                    & tf-idfs  & DAs & MCs & \begin{tabular}[c]{@{}c@{}}tf-idfs\\ +DAs\end{tabular}  &  \begin{tabular}[c]{@{}c@{}}tf-idfs\\ +MCs\end{tabular} & \begin{tabular}[c]{@{}c@{}}DA-\\ tf-idfs\end{tabular} &  \begin{tabular}[c]{@{}c@{}}MC-\\ tf-idfs\end{tabular}      \\
\hline
ag   & 0.76  & 0.60  & 0.65  & 0.75  & 0.76  & 0.77  & 0.80 \\
at   & 0.70  & 0.60  & 0.61  & 0.70  & 0.69  & 0.71  & 0.73 \\
co   & 0.75  & 0.63  & 0.64  & 0.74  & 0.75  & 0.75  & 0.80 \\
fb   & 0.75  & 0.56  & 0.63  & 0.75  & 0.76  & 0.74  & 0.76 \\
gd   & 0.74  & 0.58  & 0.61  & 0.72  & 0.77  & 0.76  & 0.73 \\
hw   & 0.66  & 0.55  & 0.61  & 0.66  & 0.70  & 0.70  & 0.68 \\
ip   & 0.54  & 0.51  & 0.55  & 0.57  & 0.53  & 0.57  & 0.58 \\
cb   & 0.73  & 0.55  & 0.69  & 0.75  & 0.75  & 0.77  & 0.77 \\
pt   & 0.69  & 0.54  & 0.66  & 0.69  & 0.73  & 0.74  & 0.75 \\
sc   & 0.73  & 0.60  & 0.61  & 0.74  & 0.76  & 0.76  & 0.76 \\
un   & 0.74  & 0.56  & 0.58  & 0.74  & 0.76  & 0.73  & 0.75 \\
\hline
avg  & 0.71  & 0.57  & 0.62  & 0.71  & 0.72  & 0.73  & \textbf{0.74}
\\
tot & 0.78  & 0.62  & 0.67  & 0.77  & 0.78  & 0.81  & \textbf{0.83}
\\
\hline
\end{tabular}}
\end{table}

\begin{table}[htb]
\caption{\label{tab:automated} F1 scores of the tasks on the automatically derived transcripts from the speech pipeline.}
\centering
\resizebox{0.42\textwidth}{!}{ 
\begin{tabular}{c|c@{\hskip 0.8cm}c@{\hskip 1.0cm}c@{\hskip 0.7cm}cc}
\hline
        & tf-idfs  & DAs & MCs & DA-tf-idfs & MC-tf-idfs\\
\hline
ag   & 0.75  & 0.62  & 0.64  & 0.77  & 0.78   \\
at   & 0.69  & 0.59  & 0.61  & 0.73  & 0.73   \\
co   & 0.73  & 0.61  & 0.66  & 0.74  & 0.75   \\
fb   & 0.75  & 0.58  & 0.64  & 0.75  & 0.77   \\
gd   & 0.68  & 0.57  & 0.60  & 0.72  & 0.70   \\
hw   & 0.65  & 0.52  & 0.63  & 0.73  & 0.69   \\
ip   & 0.55  & 0.53  & 0.53  & 0.52  & 0.53   \\
cb   & 0.71  & 0.56  & 0.63  & 0.71  & 0.75   \\
pt   & 0.66  & 0.46  & 0.64  & 0.68  & 0.72   \\
sc   & 0.72  & 0.62  & 0.63  & 0.73  & 0.76   \\
un   & 0.74  & 0.51  & 0.56  & 0.72  & 0.73   \\
\hline
avg  & 0.69  & 0.56  & 0.62  & 0.71  & \textbf{0.72} 
\\
tot & 0.76  & 0.60  & 0.66  & 0.77  & \textbf{0.80}
\\
\hline
\end{tabular}}
\end{table}

\subsection{Results of Automatically-derived Transcripts}

We next consider the end-to-end automated evaluation of CBT sessions using the transcripts generated by the speech processing pipeline described in \Cref{sec:3-1}. We perform the prediction tasks on the basic feature sets and the fused features after word augmentation. 

The experimental results are given in Table \ref{tab:automated}. Comparing the results with the ones 
in Table \ref{tab:manual}, we 
observe that while the performance of the code prediction using the automatically derived transcripts is degraded compared to evaluating on manually-derived transcripts, the drop is relatively small.
This modest performance degradation underscores both the robustness of this end-to-end speech processing system, and the room for further improvements. Again the tf-idf features achieve significantly better F1 scores than the DAs and MCs ($p<0.01$) while DAs lead to the worst performance among the basic feature sets. The DA-tf-idfs and MC-tf-idfs both outperform the tf-idfs, which is consistent with the results of the manual transcripts (Table \ref{tab:manual}). The MC-tf-idfs achieve the best overall metrics and F1 scores for the majority of the CTRS codes.

\begin{figure}[htb]
\centering
\begin{minipage}[b]{0.4\linewidth}
  \centering
  \centerline{\includegraphics[width=4cm]{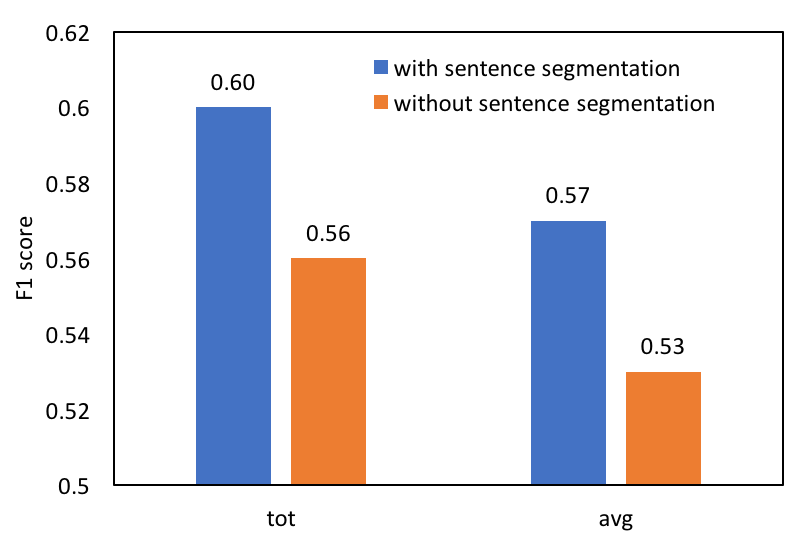}}
  \centerline{\footnotesize (a) F1 scores for the DAs}\medskip
\end{minipage}
\hspace*{2em}
\begin{minipage}[b]{0.4\linewidth}
  \centering
  \centerline{\includegraphics[width=4cm]{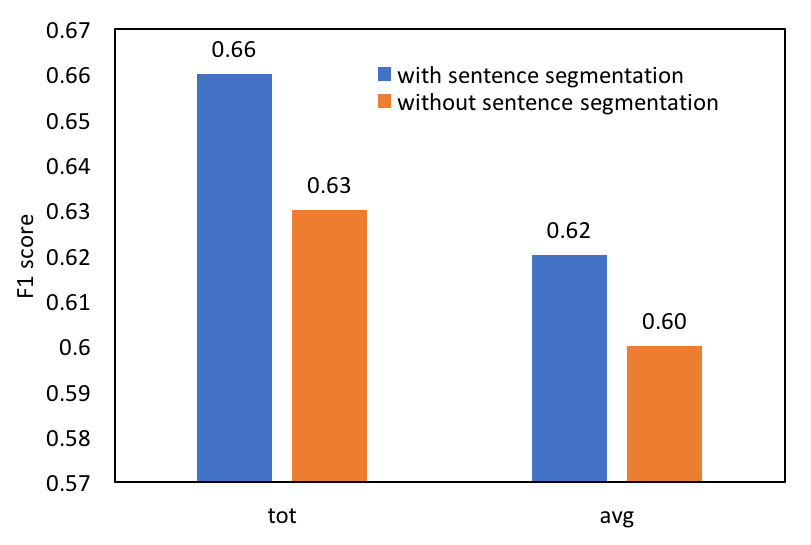}}
  \centerline{\footnotesize (b) F1 scores for the MCs}\medskip
\end{minipage}
\begin{minipage}[b]{0.4\linewidth}
  \centering
  \centerline{\includegraphics[width=4cm]{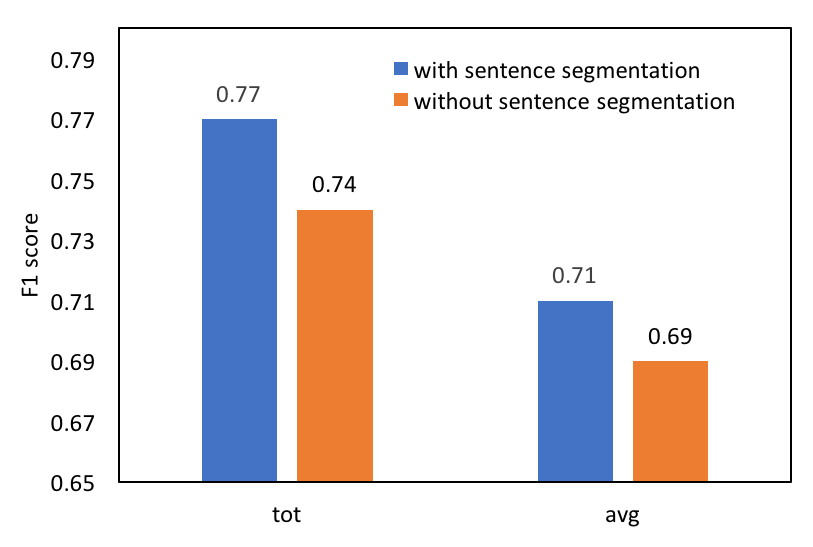}}
  \centerline{\footnotesize (c) F1 scores for the DA-tf-idfs}\medskip
\end{minipage}
\hspace*{2em}
\begin{minipage}[b]{0.4\linewidth}
  \centering
  \centerline{\includegraphics[width=4cm]{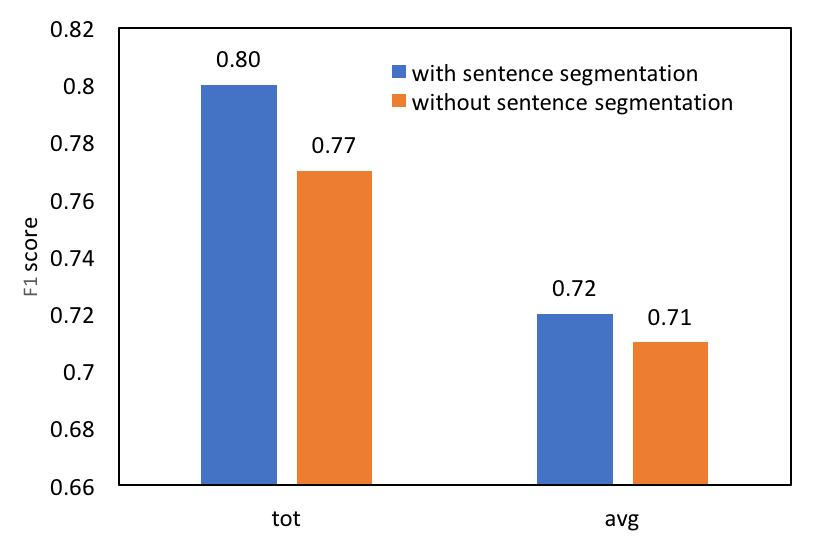}}
  \centerline{\footnotesize (d) F1 scores for the MC-tf-idfs}\medskip
\end{minipage}
\caption{Comparison of the tasks performed with and without the utterance segmentation for different feature sets.}
\label{fig:comparison}
\end{figure}

One operational challenge for utterance level processing that we often face while dealing with rich spoken interactions such as seen during therapy is the presence of multiple utterances per talk turn. This led us to investigate the role of using turn level utterance segmentation.  To 
demonstrate the effect of incorporating an utterance segmentation 
module, we experiment on the end-to-end evaluation tasks by removing this 
component from the pipeline. The comparison between the overall performances with and without the utterance segmentation is presented in Fig.~\ref{fig:comparison}. Since whether we segment the turn into sentences or not does not affect the 
output when we use individually the tf-idfs, we show the performance for the other four sets studied in Tables \ref{tab:manual} and \ref{tab:automated}. The results indicate that, for all the feature sets, 
removing the segmentation module leads to worse prediction outcomes. This confirms our hypothesis that multi-utterance turns need to be appropriately handled when we are employing utterance-specific representations such as DAs and MCs, in this study.

\section{Conclusions and Future Work}

In this paper, we employed an end-to-end approach to assess CBT psychotherapy sessions automatically without manual transcription and annotation. 
The overall CBT session quality assessment was formulated as a binary classification task, for each of the 11 target  behavioral codes, and was performed using word-level and utterance-level linguistic feature sets and their fused combinations. 
In particular, inspired by the commonality in certain elements of the therapy process between MI and CBT, we
introduce utterance-level MI codes as one of the feature sets. A new feature fusion strategy was proposed where we augmented the words of an utterance with an utterance-level tag.  
We then applied a tf-idf transformation on those augmented tokens. The experimental results showed that our end-to-end automated approach was robust 
and the final performance was comparable to using manual transcripts.
The best performance was achieved by the fused features of the tf-idfs and MI codes obtained with the new fusion strategy. Additionally, we confirmed the importance of 
including an utterance segmentation module into the pipeline. 
In the future, we will explore the importance of each code in DAs and MCs for predicting the CBT session quality and explore in further detail transfer learning between MI and CBT domains.

\section{Acknowledgements}

This work was supported by the NIH.

\bibliographystyle{IEEEtran}

\bibliography{mybib}


\end{document}